\documentclass[floatfix,nofootinbib,showpacs,showkeys,preprintnumbers]{revtex4}
\usepackage{dcolumn}
\usepackage{bm}
\usepackage[centertags]{amsmath}
\allowdisplaybreaks[1]
\usepackage{amsbsy}
\usepackage{amsfonts}
\usepackage{amssymb}
\usepackage[dvips,monochrome]{color}
\usepackage[dvips]{graphicx}
\begin{document}
\preprint{hep-ph/0107310}
\title{
Are There $\nu_\mu$ or $\nu_\tau$ in the Flux of Solar Neutrinos on Earth?
}
\author{C. Giunti}
\email{giunti@to.infn.it}
\homepage{http://www.to.infn.it/~giunti}
\affiliation{
INFN, Sezione di Torino,\\
and\\
Dipartimento di Fisica Teorica, Universit\`a di Torino,\\
Via P. Giuria 1, I--10125 Torino, Italy
}
\date{10 October 2001}
\begin{abstract}
Using the model independent method of Villante, Fiorentini, Lisi,
Fogli, Palazzo,
and the rates measured in the SNO and Super-Kamiokande
solar neutrino experiment,
we calculate the amount of active $\nu_\mu$ or $\nu_\tau$
present in the flux of solar neutrinos on Earth.
We show that the probability of $\nu_e \to \nu_{\mu,\tau}$
transitions
is larger than zero at 99.89\% CL.
We find that the averaged flux of $\nu_{\mu,\tau}$
on Earth
is larger than 0.17 times the $^8\mathrm{B}$ $\nu_e$
flux predicted by the BP2000 Standard Solar Model
at 99\% CL.
We discuss also the consequences of possible
$\nu_e\to\bar\nu_{\mu,\tau}$
or
$\nu_e\to\bar\nu_e$
transitions of solar neutrinos.
We derive a model-independent lower limit of $0.52$ at 99\% CL for
the ratio of the $^8\mathrm{B}$ $\nu_e$
flux produced in the Sun and its value in the
BP2000 Standard Solar Model.
\end{abstract}
\pacs{26.65.+t, 14.60.Pq, 14.60.Lm}
\keywords{Solar Neutrinos, Neutrino Physics, Statistical Methods}
\maketitle

The first results of the
SNO solar neutrino experiment
\cite{Ahmad:2001an}
have beautifully confirmed the existence of the solar neutrino problem.
A comparison of
the neutrino flux measured through charged-current interactions
in the SNO experiment
with the flux measured through elastic scattering interactions
in the Super-Kamiokande experiment
\cite{SK-sun-01}
shows an evidence of the presence of
active $\nu_\mu$ or\footnote{
In this paper the conjunction ``or''
is used as a logical inclusive disjunction
(the sentence is true when either or both of its constituent
propositions are true).
} $\nu_\tau$
in the solar neutrino flux
measured by the Super-Kamiokande experiment
\cite{Ahmad:2001an,Fogli:2001vr}.
Such a presence represents a very interesting
indication in favor of neutrino physics beyond the Standard Model,
most likely neutrino mixing that generates
oscillations between different flavors
(see \cite{BGG-review-98}).

The purpose of this paper is to quantify
the amount of this flux of active $\nu_\mu$ or $\nu_\tau$
in a model-independent way
in the framework of Frequentist Statistics\footnote{
Since the results that we obtain are not too close to
physical boundaries for the quantities under discussion
and we assume a normal distribution for the errors,
the numerical values in the framework of
Bayesian Probability Theory with a flat prior
are close to those obtained here,
but their meaning is different
(see, for example, Ref.~\cite{D'Agostini-99}).
}.

The authors of Refs.~\cite{Villante:1998pe,Fogli:2001nn}
have noted that the
response functions of the SNO and Super-Kamiokande (SK) experiments
to solar neutrinos
can be made approximately equal with a proper choice of the
energy thresholds of the detected electrons.
It turns out that given the threshold
$T_e^{\mathrm{SNO}} = 6.75 \, \mathrm{MeV}$,
the two response functions are approximately equal for
$T_e^{\mathrm{SK}} = 8.60 \, \mathrm{MeV}$
\cite{Fogli:2001vr}.
In this case the
SNO and Super-Kamiokande event rates normalized to the BP2000
Standard Solar Model (SSM) prediction \cite{BP2000}
can be written in a model-independent way as
\cite{Fogli:2001vr}
\begin{eqnarray}
&&
R_{\mathrm{SNO}}
=
f_{\mathrm{B}}
\,
\langle P_{\nu_e\to\nu_e} \rangle
\,,
\label{001}
\\
&&
R_{\mathrm{SK}}
=
f_{\mathrm{B}}
\,
\langle P_{\nu_e\to\nu_e} \rangle
+
f_{\mathrm{B}}
\,
\frac{\langle\sigma_{\nu_{\mu,\tau}}\rangle}{\langle\sigma_{\nu_e}\rangle}
\,
\langle P_{\nu_e\to\nu_{\mu,\tau}} \rangle
\,,
\label{002}
\end{eqnarray}
where
$f_{\mathrm{B}}$
is the ratio of the $^8\mathrm{B}$ $\nu_e$
flux produced in the Sun and its value in the SSM \cite{BP2000},
$\langle P_{\nu_e\to\nu_e} \rangle$
is the survival probability of solar $\nu_e$'s
averaged over the common SNO and Super-Kamiokande response functions,
\begin{equation}
\frac{\langle\sigma_{\nu_{\mu,\tau}}\rangle}{\langle\sigma_{\nu_e}\rangle}
=
0.152
\label{0021}
\end{equation}
is the ratio of the averaged $\nu_{\mu,\tau}$ and $\nu_e$
cross sections in the Super-Kamiokande experiment,
and
$\langle P_{\nu_e\to\nu_{\mu,\tau}} \rangle$
is the averaged probability of
$\nu_e\to\nu_{\mu,\tau}$ transitions.

Calling
\begin{equation}
R_A \equiv R_{\mathrm{SK}} - R_{\mathrm{SNO}}
\,,
\label{005}
\end{equation}
from Eqs.(\ref{001}) and (\ref{002}) we have
\begin{equation}
R_A
=
f_{\mathrm{B}}
\,
\frac{\langle\sigma_{\nu_{\mu,\tau}}\rangle}{\langle\sigma_{\nu_e}\rangle}
\,
\langle P_{\nu_e\to\nu_{\mu,\tau}} \rangle
\,.
\label{006}
\end{equation}
Therefore,
$R_A$
is the rate of $\nu_{\mu,\tau}$-induced events
in the Super-Kamiokande experiment,
relative to the $\nu_e$-induced rate predicted by the SSM.

Considering the data of the Super-Kamiokande
experiment above the energy threshold
$T_e^{\mathrm{SK}} = 8.60 \, \mathrm{MeV}$
and
the BP2000 Standard Solar Model \cite{BP2000},
the measured values of
$R_{\mathrm{SNO}}$
and
$R_{\mathrm{SK}}$
are:
\begin{eqnarray}
&&
R_{\mathrm{SNO}}^{\mathrm{exp}}
=
0.347 \pm 0.029
\quad
\text{\protect\cite{Ahmad:2001an}}
\,,
\label{003}
\\
&&
R_{\mathrm{SK}}^{\mathrm{exp}}
=
0.451 \pm 0.017
\quad
\text{\protect\cite{SK-sun-01,Fogli:2001vr}}
\,.
\label{004}
\end{eqnarray}
Adding in quadrature the uncertainties of
$R_{\mathrm{SNO}}$
and
$R_{\mathrm{SK}}$,
for
$R_A$
we obtain:
\begin{equation}
R_A^{\mathrm{exp}}
=
0.104 \pm 0.034
\,.
\label{007}
\end{equation}
The standard deviation of $R_A^{\mathrm{exp}}$
is
\begin{equation}
\sigma_A^{\mathrm{exp}} = 0.034
\,,
\label{008}
\end{equation}
and we have
\begin{equation}
\frac{ R_A^{\mathrm{exp}} }{ \sigma_A^{\mathrm{exp}} }
=
3.06 \pm 1
\,.
\label{009}
\end{equation}
Hence,
the central value of
$R_A$
is $3.06\sigma$ away from zero,
implying an evidence of solar
$\nu_e\to\nu_{\mu,\tau}$ transitions \cite{Ahmad:2001an,Fogli:2001vr}.
Our purpose is to quantify the probability
of these transitions
and possibly derive a lower limit.

The authors of Ref.~\cite{Ahmad:2001an}
calculate the probability of a fluctuation
larger than the observed one assuming $R_A=0$:
for normally distributed errors
the probability of a fluctuation larger than
3.06$\sigma$
from the mean
is
0.11\%.

Recently some frequentist methods have been proposed
that allow to obtain always meaningful confidence intervals
with correct coverage
for quantities like $R_A$
that are bound to be positive by definition
\cite{Feldman-Cousins-98,Ciampolillo-98,%
Giunti-bo-99,
Mandelkern-Schultz-99}.
In particular,
the Unified Approach proposed in Ref.~\cite{Feldman-Cousins-98}
has been widely publicized by
the Particle Data Group \cite{PDG}
and used by several experimental collaborations.

\begin{figure}[p]
\begin{center}
\includegraphics[bb=80 420 405 755, width=0.5\textwidth]{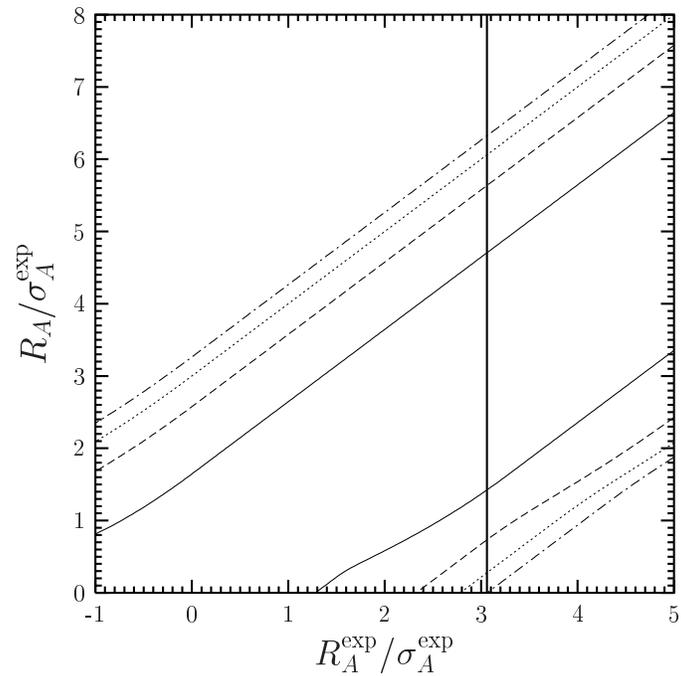}
\end{center}
\caption{ \label{uabelt}
Confidence belts in the Unified Approach
\protect\cite{Feldman-Cousins-98}
for a normal distribution with unit standard deviation.
The regions between the
solid, long-dashed, dotted and dash-dotted lines
correspond, respectively,
to
90\% ($1.64\sigma$),
99\% ($2.58\sigma$),
99.73\% ($3\sigma$) and
99.89\% ($3.06\sigma$) CL.
The thick solid vertical line represent the measured
value of
$ R_A^{\mathrm{exp}} / \sigma_A^{\mathrm{exp}} $
(Eq.~(\ref{009})).
}
\end{figure}

Using the Unified Approach
we can derive confidence intervals for $R_A$.
Figure~\ref{uabelt}
shows the confidence belts in the Unified Approach
for a normal distribution with unit standard deviation
for
90\% ($1.64\sigma$),
99\% ($2.58\sigma$),
99.73\% ($3\sigma$) and
99.89\% ($3.06\sigma$) CL.
One can see that the measured value (\ref{009})
of
$ R_A^{\mathrm{exp}} / \sigma_A^{\mathrm{exp}} $
implies that
\begin{equation}
0
<
\frac{ R_A }{ \sigma_A^{\mathrm{exp}} }
<
6.32
\quad
\mbox{at}
\quad
99.89\% \, \mathrm{CL}
\,,
\label{0101}
\end{equation}
\textit{i.e.}
active $\nu_\mu$ or $\nu_\tau$
are present in the solar neutrino flux on Earth at
99.89\% CL.
Equation (\ref{0101}) implies that there is a
0.11\% probability that
the true value of $R_A / \sigma_A^{\mathrm{exp}}$
is zero or larger than 6.32.
This probability is the same as the
probability of a fluctuation larger than
3.06$\sigma$
calculated in Ref.~\cite{Ahmad:2001an}
assuming $R_A=0$.
However,
our result have been derived without making any assumption
on the true unknown value of $R_A$
and has a well defined meaning in the
framework of Frequentist Statistics:
whatever the true value of $R_A$,
the interval (\ref{0101}) belongs to a set of
intervals that could be obtained in the same way from repeated measurements
and have the property that
99.89\% of these intervals cover the true value of
$R_A / \sigma_A^{\mathrm{exp}}$.

In order to derive a lower limit for the
averaged flux of $\nu_{\mu,\tau}$ on Earth,
we consider in the following 99\% confidence intervals.
From Fig.~\ref{uabelt}
we obtain
\begin{equation}
0.74
<
\frac{ R_A }{ \sigma_A^{\mathrm{exp}} }
<
5.63
\qquad
(99\% \, \mathrm{CL})
\,,
\label{010}
\end{equation}
whose meaning is that
there is a 99\% probability
that the interval (\ref{010})
covers the true unknown value of $R_A / \sigma_A^{\mathrm{exp}}$.

For
$
f_{\mathrm{B}}
\,
\langle P_{\nu_e\to\nu_{\mu,\tau}} \rangle
$,
that gives the flux of active $\nu_{\mu,\tau}$
averaged over the common Super-Kamiokande and SNO response function,
relative to the SSM $^{8}\mathrm{B}$ $\nu_e$ flux,
we find
\begin{equation}
0.17
<
f_{\mathrm{B}}
\,
\langle P_{\nu_e\to\nu_{\mu,\tau}} \rangle
<
1.26
\qquad
(99\% \, \mathrm{CL})
\,.
\label{011}
\end{equation}
Hence,
we can say that the averaged flux of $\nu_{\mu,\tau}$
on Earth
is larger than 0.17 times the $^8\mathrm{B}$ $\nu_e$
flux predicted by the Standard Solar Model
at 99\% CL.
This is an evidence in favor of relatively large
$\nu_e\to\nu_{\mu,\tau}$ transitions
if $f_B$ is not too large.

One could argue that it is possible to derive a more stringent
lower limit for
$
f_{\mathrm{B}}
\,
\langle P_{\nu_e\to\nu_{\mu,\tau}} \rangle
$
by calculating a confidence belt without left edge,
instead of the one in Figure~\ref{uabelt}
calculated in the Unified Approach.
Such a procedure is not acceptable,
because it would lead to undercoverage
if not chosen a priori,
independently from the data,
as shown in Ref.~\cite{Feldman-Cousins-98}
for the case of upper limits.
The correct procedure is to choose a priori a method
like the Unified Approach that gives always sensible results
and apply it to the data,
as we have done here.
A priori one could have chosen another method,
as those presented in Refs.~\cite{Ciampolillo-98,%
Giunti-bo-99,
Mandelkern-Schultz-99},
that may have even better properties than the Unified Approach
\cite{Giunti-Laveder-physical-00,Giunti-Laveder-power-00},
but we have verified that the intervals (\ref{0101})--(\ref{011})
do not change significantly.

Unfortunately,
we cannot derive a model independent lower limit for
the averaged
$\nu_e\to\nu_{\mu,\tau}$
probability
$\langle P_{\nu_e\to\nu_{\mu,\tau}} \rangle$,
because $f_{\mathrm{B}}$ could be large.
However,
from Figure~\ref{uabelt}
we can say that
$ R_A / \sigma_A^{\mathrm{exp}} > 0 $
at
99.89\% CL
(see Eq.~(\ref{0101})),
and hence
\begin{equation}
P_{\nu_e\to\nu_{\mu,\tau}} > 0
\quad
\mbox{at}
\quad
99.89\% \, \mathrm{CL}
\label{0111}
\end{equation}
in the range of neutrino energies covered by
the common SNO and Super-Kamiokande
response function presented in Ref.~\cite{Fogli:2001vr}.

On the other hand,
it is interesting to note that the relations (\ref{001}) and (\ref{002})
allow to derive a model-independent lower limit
for $f_{\mathrm{B}}$,
taking into account that
\begin{equation}
\langle P_{\nu_e\to\nu_{\mu,\tau}} \rangle
\leq
1 - \langle P_{\nu_e\to\nu_e} \rangle
\,.
\label{012}
\end{equation}
Using this inequality,
from Eqs.~(\ref{001}) and (\ref{002})
we obtain
\begin{equation}
f_{\mathrm{B}}
\geq
\frac{\langle\sigma_{\nu_e}\rangle}{\langle\sigma_{\nu_{\mu,\tau}}\rangle}
\,
R_{\mathrm{SK}}
-
\left(
\frac{\langle\sigma_{\nu_e}\rangle}{\langle\sigma_{\nu_{\mu,\tau}}\rangle}
-
1
\right)
R_{\mathrm{SNO}}
\equiv
f_{\mathrm{B,min}}
\,.
\label{013}
\end{equation}
From Eqs.~(\ref{0021}), (\ref{003}) and (\ref{004}),
the experimental value of $f_{\mathrm{B,min}}$
is
\begin{equation}
f_{\mathrm{B,min}}^{\mathrm{exp}}
=
1.031 \pm 0.197
\,.
\label{014}
\end{equation}
Since the central value of $f_{\mathrm{B,min}}^{\mathrm{exp}}$
is $5.2\sigma$ away from zero,
we can calculate the resulting 99\% CL interval for
$f_{\mathrm{B,min}}$
using the Central Intervals method (see \cite{PDG}),
that gives the same result as the Unified Approach
far from the physical boundary $f_{\mathrm{B,min}} > 0$.
Since in the Central Intervals method 99\% CL
corresponds to $2.58\sigma$,
we obtain the confidence interval
\begin{equation}
0.52
<
f_{\mathrm{B,min}}
<
1.54
\qquad
(99\% \, \mathrm{CL})
\,.
\label{015}
\end{equation}
Therefore,
we can conclude that the SNO and Super-Kamiokande data imply
the model-independent lower limit
\begin{equation}
f_{\mathrm{B}}
>
0.52
\qquad
(99\% \, \mathrm{CL})
\,.
\label{016}
\end{equation}
This is a very interesting information for the physics of the Sun.

So far we have not considered the possible
existence of exotic mechanisms
that produce
$\nu_e\to\bar\nu_{\mu,\tau}$
or
$\nu_e\to\bar\nu_e$
transitions\footnote{
We would like to thank a referee of the first version of this paper
for pointing out the possibility of
$\nu_e\to\bar\nu_{\mu,\tau}$
transitions of solar neutrinos.
}
(in addition or alternative to $\nu_e\to\nu_{\mu,\tau}$ transitions),
such as resonant spin-flavor precession
of Majorana neutrinos\footnote{
In the case of Majorana neutrinos the right-handed states
are conventionally called antineutrinos.
} \cite{Lim:1988tk,Akhmedov:1988uk}.
In this case,
Eq.~(\ref{002})
must be replaced with
\begin{equation}
R_{\mathrm{SK}}
=
f_{\mathrm{B}}
\,
\langle P_{\nu_e\to\nu_e} \rangle
+
f_{\mathrm{B}}
\left[
\frac{\langle\sigma_{\nu_{\mu,\tau}}\rangle}{\langle\sigma_{\nu_e}\rangle}
\,
\langle P_{\nu_e\to\nu_{\mu,\tau}} \rangle
+
\frac{\langle\sigma_{\bar\nu_{\mu,\tau}}\rangle}{\langle\sigma_{\nu_e}\rangle}
\,
\langle P_{\nu_e\to\bar\nu_{\mu,\tau}} \rangle
+
\frac{\langle\sigma_{\bar\nu_e}\rangle}{\langle\sigma_{\nu_e}\rangle}
\,
\langle P_{\nu_e\to\bar\nu_e} \rangle
\right]
\,,
\label{021}
\end{equation}
and
Eq.~(\ref{006}) with
\begin{equation}
R_A
=
f_{\mathrm{B}}
\left[
\frac{\langle\sigma_{\nu_{\mu,\tau}}\rangle}{\langle\sigma_{\nu_e}\rangle}
\,
\langle P_{\nu_e\to\nu_{\mu,\tau}} \rangle
+
\frac{\langle\sigma_{\bar\nu_{\mu,\tau}}\rangle}{\langle\sigma_{\nu_e}\rangle}
\,
\langle P_{\nu_e\to\bar\nu_{\mu,\tau}} \rangle
+
\frac{\langle\sigma_{\bar\nu_e}\rangle}{\langle\sigma_{\nu_e}\rangle}
\,
\langle P_{\nu_e\to\bar\nu_e} \rangle
\right]
\,.
\label{022}
\end{equation}
Using the
$^8\mathrm{B}$ neutrino spectrum
given in Ref.~\cite{Bahcall:1996qv},
the neutrino-electron elastic scattering cross section calculated in
Ref.~\cite{Bahcall:1995mm} taking into account radiative corrections,
and the Super-Kamiokande energy resolution
given in Ref.~\cite{Nakahata:1998pz},
we obtain
the following values for the ratios of the averaged
cross sections in the Super-Kamiokande experiment
for the threshold energy
$T_e^{\mathrm{SK}} = 8.60 \, \mathrm{MeV}$:
\begin{equation}
\frac{\langle\sigma_{\bar\nu_{\mu,\tau}}\rangle}{\langle\sigma_{\nu_e}\rangle}
=
0.114
\,,
\qquad
\frac{\langle\sigma_{\bar\nu_e}\rangle}{\langle\sigma_{\nu_e}\rangle}
=
0.120
\,.
\label{023}
\end{equation}
Hence,
we have the useful inequalities
\begin{equation}
\frac{\langle\sigma_{\bar\nu_{\mu,\tau}}\rangle}{\langle\sigma_{\nu_e}\rangle}
<
\frac{\langle\sigma_{\bar\nu_e}\rangle}{\langle\sigma_{\nu_e}\rangle}
<
\frac{\langle\sigma_{\nu_{\mu,\tau}}\rangle}{\langle\sigma_{\nu_e}\rangle}
\,.
\label{025}
\end{equation}

The lower bound in Eq.~(\ref{0101})
implies the existence of solar
$\nu_e\to\nu_{\mu,\tau}$
or
$\nu_e\to\bar\nu_{\mu,\tau}$
or
$\nu_e\to\bar\nu_e$
transitions
at 99.89\% CL.
The inequalities in Eq.~(\ref{010}) imply that
the quantity on the right-hand side of Eq.~(\ref{022})
is limited in the interval
$(0.025,0.19)$ at 99\% CL.
Using the inequalities (\ref{025}), we obtain
\begin{equation}
0.17
<
f_{\mathrm{B}}
\left[
\langle P_{\nu_e\to\nu_{\mu,\tau}} \rangle
+
\langle P_{\nu_e\to\bar\nu_{\mu,\tau}} \rangle
+
\langle P_{\nu_e\to\bar\nu_e} \rangle
\right]
<
1.67
\qquad
(99\% \, \mathrm{CL})
\,.
\label{031}
\end{equation}
Therefore,
the averaged flux of
$\nu_{\mu}$,
$\nu_{\tau}$,
$\bar\nu_{\mu}$,
$\bar\nu_{\tau}$ and
$\bar\nu_{e}$
on Earth
is larger than 0.17 times the $^8\mathrm{B}$ $\nu_e$
flux predicted by the BP2000 Standard Solar Model
at 99\% CL.

Let us derive now the most general model-independent lower limit
for $f_{\mathrm{B}}$
(assuming only that the Super-Kamiokande and SNO events
are produced by neutrinos or antineutrinos
generated as $\nu_e$ from $^8\mathrm{B}$
decay in the Sun).
Using the inequality
\begin{equation}
\langle P_{\nu_e\to\nu_{\mu,\tau}} \rangle
+
\langle P_{\nu_e\to\bar\nu_{\mu,\tau}} \rangle
+
\langle P_{\nu_e\to\bar\nu_e} \rangle
\leq
1 - \langle P_{\nu_e\to\nu_e} \rangle
\label{024}
\end{equation}
and those in Eq.~(\ref{025}),
from Eqs.~(\ref{001}) and (\ref{021})
we obtain again the limit in Eq.~(\ref{013}).
Therefore,
Eq.~(\ref{016}) gives
the most general model-independent lower limit
for $f_{\mathrm{B}}$ following from the SNO and Super-Kamiokande data.

In conclusion,
we have considered the model independent relations
(\ref{001}), (\ref{002})
\cite{Villante:1998pe,Fogli:2001nn,Fogli:2001vr}
(and (\ref{001}), (\ref{021}))
and the rates measured in the
SNO \cite{Ahmad:2001an} and Super-Kamiokande \cite{SK-sun-01}
solar neutrino experiment
in the framework of Frequentist Statistics.
We have shown that
the probability
of $\nu_e \to \nu_{\mu,\tau}$
(and
$\nu_e \to \bar\nu_{\mu,\tau}$,
$\nu_e \to \bar\nu_e$)
transitions
is larger than zero at 99.89\% CL
in the range of neutrino energies covered by
the common SNO and Super-Kamiokande
response function.
We have found that
the flux of $\nu_{\mu,\tau}$
(and
$\bar\nu_{\mu,\tau}$,
$\bar\nu_e$)
on Earth
averaged over the common SNO and Super-Kamiokande
response functions
is larger than 0.17 times the $^8\mathrm{B}$ $\nu_e$
flux predicted by the BP2000 Standard Solar Model
at 99\% CL.
We have derived a model-independent lower limit of $0.52$
at 99\% CL
for
the ratio $f_{\mathrm{B}}$ of the $^8\mathrm{B}$ $\nu_e$
flux produced in the Sun and its value in the
BP2000 Standard Solar Model \cite{BP2000}.

\end{document}